\documentstyle[prl,aps,epsfig,floats]{revtex}

\newcommand\beq{\begin{equation}}
\newcommand\eeq{\end{equation}}
\newcommand\bea{\begin{eqnarray}}
\newcommand\eea{\end{eqnarray}}
\newcommand\non{\nonumber}

\newcommand\bib{\bibitem}

\begin{document}

\draft

\textheight=23.8cm
\twocolumn[\hsize\textwidth\columnwidth\hsize\csname@twocolumnfalse\endcsname

\title{\Large Magnetic properties of a helical spin chain with alternating
isotropic and anisotropic spins: magnetization plateaus and finite entropy}
\author{\bf V. Ravi Chandra$^1$, S. Ramasesha$^2$, and Diptiman Sen$^1$} 
\address{\it $^1$ Centre for Theoretical Studies, Indian Institute of Science,
Bangalore 560012, India \\
\it $^2$ Solid State and Structural Chemistry Unit, Indian Institute of 
Science, Bangalore 560012, India}

\maketitle

\begin{abstract}

We study a model which could explain some of the unusual magnetic properties 
observed for the one-dimensional helical spin system $Co(hfac)_2 NITPhOMe$. 
One of the properties observed is that the magnetization shows plateaus near 
zero and near one-third of the saturation value if a magnetic field is applied
along the helical axis, but not if the field is applied in the plane 
perpendicular to that axis. The system consists of a spin-1/2 chain in which 
cobalt ions (which are highly anisotropic with an easy axis ${\bf e}_i$) and 
organic radicals (which are isotropic) alternate with each other. The easy axis
of the cobalts ${\bf e}_i$ lie at an angle $\theta_i$ with respect to the 
helical axis, while the projection of ${\bf e}_{i+1} - {\bf e}_i$ on the plane
perpendicular to the helical axis is given by $2\pi /3$. For temperatures and 
magnetic fields which are much smaller than the coupling between the 
nearest-neighbor cobalts and radicals, one can integrate out the radicals to 
obtain an Ising model for the cobalts; this enables one to compute the 
thermodynamic properties of the system using the transfer matrix approach. We 
consider a model in which the tilt angles $\theta_i$ are allowed to vary with 
$i$ with period three; we find that for certain patterns of $\theta_i$, the 
system shows the magnetization plateaus mentioned above. At the ends of the 
plateaus, the entropy is finite even at very low temperatures, while the 
magnetic susceptibility and specific heat also show some interesting features.

\end{abstract}
\vskip .6 true cm

\pacs{~~ PACS numbers: ~75.10.Pq, ~05.50.+q, ~75.50.Xx}
\vskip.5pc
]
\vskip .6 true cm

The last several years have witnessed extensive studies of one-dimensional 
systems and molecular clusters with a variety of interesting magnetic 
properties, both static and dynamic \cite{miller}. 
Very recently, there have been some experimental studies of a one-dimensional 
molecular system $Co(hfac)_2NITPhOMe$ (to be 
called $CoPhOMe$ henceforth) which shows some unusual behavior in the presence
of a time-dependent magnetic field \cite{caneschi1,caneschi2}. The
system has a helical structure, in which cobalt ions and organic radicals
(all carrying spin-1/2) alternate, with a repeat period of three cobalts for
every turn of the helix; this is shown in Fig. 1. Below a certain temperature,
the time scale associated with the variation of the magnetization is found to 
become extremely long (leading to a pronounced hysteresis) if 
the magnetic field is applied along the helical axis (called the $c$ axis), 
but not if the field is in the plane perpendicular to that axis (called the 
$a-b$ plane). It is also found that the magnetization shows some plateaus 
(which become more pronounced at lower temperatures) if the magnetic field 
points along the $c$ axis, but not if it is in the $a-b$ plane.

\begin{figure}[htb]
\begin{center}
\epsfig{figure=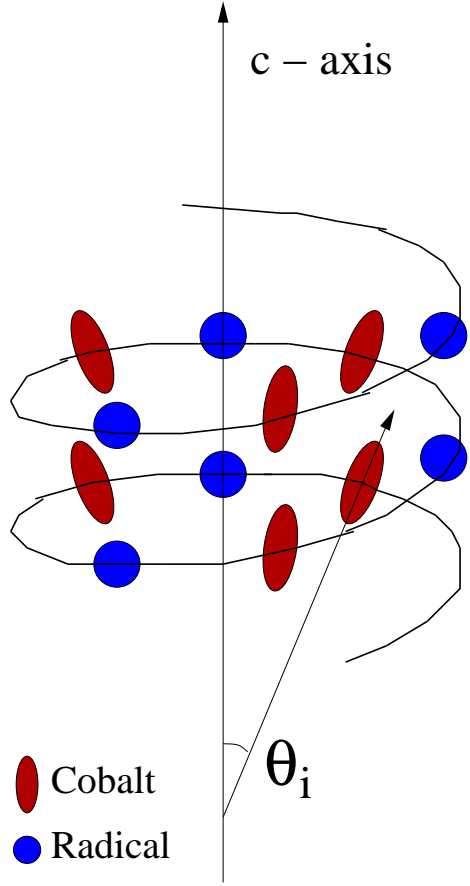,width=7cm}
\end{center}
\caption{The structure of the molecular chain $CoPhOMe$. The cobalt spins are
anisotropic with a local axis denoted by ${\bf e}_i$ which is tilted by an 
angle $\theta_i$ with respect to the helical axis $c$. The angle between the 
projections of ${\bf e}_{i+1}$ and ${\bf e}_i$ on the $a-b$ plane is equal 
to $2\pi /3$. The organic radical spins are isotropic.}
\end{figure}

In this work, we will consider the second feature mentioned above, namely, the
appearance of some plateaus with non-trivial magnetizations when the magnetic 
field is applied along the $c$ axis. We will present a model which can 
qualitatively explain this feature. Our model is a variation of the one 
considered in the earlier studies of this system 
\cite{caneschi1,caneschi2,vindigni}; for reasons explained below, the model 
presented in those papers is not able to explain the magnetization plateaus.

We begin by presenting the model introduced for this system in the earlier 
papers \cite{caneschi1,caneschi2,vindigni}. Both the cobalt ions and the 
organic radicals carry spin-1/2. The cobalt ions are highly anisotropic; they 
have an easy axis ${\bf e}_i$ which is tilted by an angle $\theta_i$ with 
respect to the $c$ axis. Further, the projection of ${\bf e}_{i+1} - {\bf e}_i$
on the $a-b$ plane is given by $2 \pi /3$. (We assume that ${\bf e}_i$ is a 
unit vector). If we identify the $c$ axis with the $\bf z$ axis, the three 
components of ${\bf e}_i$ are given by $(\sin \theta_i \cos 2\pi (i-1)/3 , 
\sin \theta_i \sin 2\pi (i-1)/3 , \cos \theta_i )$. Due to the anisotropy, the 
cobalt spins can be described classically using Ising variables $\sigma_i$. The
organic radicals are completely isotropic, and their spins have to be treated 
quantum mechanically. The earlier papers assumed the tilt angles $\theta_i$ to 
be the same for all the cobalts. However, we will allow the $\theta_i$ to vary 
with $i$ (but with a period of three keeping the pitch of the helix the same 
as in the earlier models); as we will see, this variation seems to be 
necessary in order to reproduce the observed magnetization plateaus.

In the $i^{\rm th}$ cobalt-radical pair, let us denote the component of the 
cobalt spin along its easy axis by $\sigma_i$ 
(where $\sigma_i = \pm 1$), and the spin operators of the radical by 
${\bf T}_i$ (these are given by half the Pauli matrices). In the presence of 
a magnetic field $\bf B$, the Hamiltonian for this system is given by 
\bea
H_{CR} ~=~ \sum_i ~[ & & \frac{J}{2} ~\sigma_i {\bf e}_i \cdot (~{\bf T}_i +
{\bf T}_{i-1} ~) ~ \non \\
& & -~ \mu_B {\bf B} \cdot (~ \frac{1}{2} g_C \sigma_i {\bf e}_i + g_R 
{\bf T}_i ~) ~]~,
\label{hcr}
\eea
where $g_C$ and $g_R$ denote the gyromagnetic ratios of the cobalt and
radical spins respectively, and $\mu_B = e\hbar /(2mc)$ is the Bohr magneton
(we note that $\mu_B /k_B = 0.672$ K/Tesla). Fits to the magnetization data at 
different temperatures seem to lead to somewhat different values of the various
parameters. One set of parameters which has been quoted in some of the papers 
is as follows: $J/k_B \sim 400$ K (antiferromagnetic in sign), $g_C = 9$, and 
$g_R =2$, and the tilt angle $\theta$ is in the vicinity of the magic angle 
$\theta_0 = \cos^{-1} (1/{\sqrt 3}) \simeq 54.74^o$ \cite{caneschi2,vindigni}.
(Large values of the effective $g$ factor given by $g_{\cal J} {\cal J}$ are 
known to arise in high spin systems when a strong uniaxial anisotropy restricts
the accessible spin states ${\cal J}_z$ to $\pm {\cal J}$ at low temperatures 
\cite{fukazawa,rosenkranz}).

The data which indicates magnetization plateaus lies at a temperature of 
about $2$ K and a magnetic field of up to $3$ Tesla. Since these 
temperatures and magnetic fields are much smaller than the value of $J/k_B$ and
$J/\mu_B$ respectively, we will make the approximation from that each 
radical spin is aligned in a direction which is entirely dictated by the
directions of its two neighboring cobalt spins. Namely, we assume that the 
expectation value of ${\bf T}_i$ is given by
\beq
\langle {\bf T}_i \rangle ~=~ - ~\frac{1}{2} ~\frac{{\sigma_i \bf e}_i ~+~ 
\sigma_{i+1} {\bf e}_{i+1}}{\sqrt {2 ~+~ 2 \sigma_i \sigma_{i+1} ~{\bf e}_i 
\cdot {\bf e}_{i+1}}}~.
\label{ti}
\eeq
Upon substituting this in Eq. (\ref{hcr}), we obtain an effective Hamiltonian
defined purely in terms of the cobalt Ising variables $\sigma_i$,
\bea
H_{1C} = \sum_i & & [~ - \frac{J}{4} ~{\sqrt {2 + 2 \sigma_i \sigma_{i+1} ~
{\bf e}_i \cdot {\bf e}_{i+1}}} \non \\
& & - \frac{\mu_B}{2} ~{\bf B} \cdot {\bf e}_i ~\sigma_i ~(~ g_C - 
\frac{g_R}{\sqrt {2 + 2 \sigma_i \sigma_{i+1} ~
{\bf e}_i \cdot {\bf e}_{i+1}}} \non \\
& & \quad \quad \quad \quad \quad \quad \quad - \frac{g_R}{\sqrt {2 + 
2 \sigma_i \sigma_{i-1} ~{\bf e}_i \cdot {\bf e}_{i-1}}} ~) ~] ~. \non \\
& & 
\label{h1c}
\eea

As mentioned above, the experimental data indicates that the tilt angles 
$\theta_i$ are close to the magic angle $\theta_0$. If all the $\theta_i$
were {\it exactly} equal to $\theta_0$, we would have ${\bf e}_i \cdot
{\bf e}_{i+1} = 0$. Then the Hamiltonian in Eq. (\ref{h1c}) would have no 
interactions between neighboring cobalts, and the (subtracted) two-spin 
correlations, $\langle (\sigma_i - \langle \sigma_i \rangle) (\sigma_j - 
\langle \sigma_j \rangle) \rangle$, would be 
strictly zero for $i \ne j$ at any temperature. [This is called a disorder
point; it corresponds to the smaller eigenvalue of the transfer matrix 
(discussed below) being equal to zero \cite{vindigni,selke}].

Motivated by the ranges of the various experimental parameters, let us assume 
that $\delta \theta_i \equiv \theta_i - \theta_0$ are small numbers (in 
radians), so that 
\beq
{\bf e}_i \cdot {\bf e}_{i+1} ~\simeq ~-~ \frac{1}{\sqrt 2} ~(~ \delta 
\theta_i ~+~ \delta \theta_{i+1} ~)~
\eeq
is much less than 1 in magnitude. We also assume that $J (\delta \theta_i + 
\delta \theta_{i+1})$ is of the same order as (or larger than) than the 
magnitude of $\mu_B | {\bf B} |$. Then Eq. (\ref{h1c}) can be approximately 
written, up to a constant, as
\bea
H_{2C} &=& \sum_i ~[~ J_{i,i+1} ~\sigma_i \sigma_{i+1} ~-~ \frac{\mu_B}{2} ~
g_{eff} ~ {\bf B} \cdot {\bf e}_i^0 ~ \sigma_i ~] ~, \non \\
J_{i,i+1} &=& \frac{J}{8} ~(~ \delta \theta_i ~+~ \delta 
\theta_{i+1} ~)~, \non \\
g_{eff} &=& g_C ~-~ {\sqrt 2} ~ g_R ~,
\label{h2c}
\eea
and 
\bea
{\bf e}_1^0 &=& (~{\sqrt {2/3}}~,~ 0~,~ 1/{\sqrt 3} ~)~, \non \\
{\bf e}_2^0 &=& (~-~ 1/{\sqrt 6}~,~ 1/{\sqrt 2}~,~ 1/{\sqrt 3} ~)~, \non \\
{\bf e}_3^0 &=& (~-~ 1/{\sqrt 6}~,~ - ~1/{\sqrt 2}~,~ 1/{\sqrt 3}~) ~.
\eea
The effective nearest neighbor Ising interaction $J_{i,i+1}$ in Eq. (\ref{h2c})
is ferromagnetic or antiferromagnetic depending on whether $\delta \theta_i + 
\delta \theta_{i+1}$ is negative or positive.

In the earlier papers \cite{caneschi1,caneschi2,vindigni}, $\theta_i$ had been 
assumed to take the same value $\theta$ for all $i$. Then the effective Ising
interaction is given by 
\beq
J_{i,i+1} ~=~ \frac{J}{4} ~\delta \theta ~.
\label{jeff}
\eeq
The thermodynamic properties of this system model can be calculated easily 
using the transfer matrix method. If $g_{eff} > 0$, and the magnetic field is 
large compared to $J \delta \theta$ (but much smaller than $J$), then Eq. 
(\ref{h2c}) implies that the magnetization per cobalt-radical pair will take 
a value given by
\bea
M_S &=& \frac{\mu_B}{6} ~g_{eff} ~ \sum_{i=1}^3 ~|~ {\hat B} \cdot 
{\bf e}_i^0 ~|~, \non \\
{\hat B} &=& \frac{\bf B}{| {\bf B} |} ~.
\label{mag}
\eea
We will henceforth refer to $M_S$ as the saturation magnetization. [If the 
magnetic field becomes much larger than $J/\mu_B$, i.e., about 600 Tesla, then
the original Hamiltonian in Eq. (\ref{hcr}) implies that the magnetization will
reach the final saturation value of $(\mu_B /6) (g_C \sum_i |{\hat B} \cdot 
{\bf e}_i^0 | + 3 g_R)$. But this kind of field strength is not experimentally
accessible at present]. Let us now consider the case of very low temperatures 
and a magnetic field applied along the $c$ axis; then all the cobalt spins 
experience the same magnetic field strength ${\bf B} \cdot {\bf e}_i^0 = | 
{\bf B} | /{\sqrt 3}$. The magnetization will show a plateau at $M=0$ if the 
effective interaction in Eq. (\ref{jeff}) is positive (antiferromagnetic), but
not if it is negative (ferromagnetic). For large fields, the magnetization will
saturate at $M=M_S= \mu_B g_{eff} /(2{\sqrt 3})$. We thus see that there is no
magnetization plateau at fractional values of $M_S$ (such as $M_S /3$) 
regardless of what the sign of $\delta \theta$ in Eq. (\ref{jeff}) is; namely,
states with magnetization equal to $M_S /3$ are not the lowest energy states 
for any value of the field.

We therefore require a slightly different model in order to obtain 
magnetization plateaus at both $M=0$ and $M=M_S /3$ as the experimental
data seems to suggest \cite{caneschi2}. After considering 
several possible variations of the basic model, we have found that the 
following idea works. We assume that all the cobalt spins in a single
molecular chain do not have the same angle of tilt with respect to the
$c$ axis. We further assume that the angles $\theta_i$ take three
different values $\theta_1, \theta_2$ and $\theta_3$ for three successive
cobalts, and that they repeat periodically thereafter. The repeat period of
three makes it plausible that there could be a magnetization
plateau at $M_S /3$ (corresponding to a state with $\sigma_i$ repeating 
as $1,1,-1$, i.e., a $\uparrow \uparrow \downarrow$ spin alignment). 
However, it turns out that $\theta_1, \theta_2$ and 
$\theta_3$ need to satisfy some additional conditions as we will now discuss.

Since the $\theta_i$'s repeat with period three, the thermodynamic properties 
of the system can again be found using the transfer matrix method. If the 
number of cobalt-radical pairs is denoted by $N$, and we use periodic boundary
conditions (taking $N$ to be a multiple of 3), then the partition function can
be written as
\beq
Z ~=~ {\rm Tr} ~(~ A_1 ~A_2 ~A_3 ~)^{N/3} ~,
\label{part}
\eeq
where the matrix elements of the $2 \times 2$ matrices $A_i$ are given by 
\bea
(A_i)_{11} &=& \exp ~[ -\beta J_{i,i+1} + ~\frac{\beta \mu_B g_{eff}}{4}~
{\bf B} \cdot ({\bf e}_i^0 + {\bf e}_{i+1}^0) ], \non \\
(A_i)_{12} &=& \exp ~[ \beta J_{i,i+1} + ~\frac{\beta \mu_B g_{eff}}{4}~
{\bf B} \cdot ({\bf e}_i^0 - {\bf e}_{i+1}^0) ], \non \\
(A_i)_{21} &=& \exp ~[ \beta J_{i,i+1} + ~\frac{\beta \mu_B g_{eff}}{4}~
{\bf B} \cdot (- {\bf e}_i^0 + {\bf e}_{i+1}^0) ], \non \\
(A_i)_{22} &=& \exp ~[ -\beta J_{i,i+1} + ~\frac{\beta \mu_B g_{eff}}{4}~
{\bf B} \cdot (- {\bf e}_i^0 - {\bf e}_{i+1}^0) ], \non \\
& &
\label{ai}
\eea
with $\beta = 1/(k_B T)$. The magnetization per cobalt-radical pair is then 
given by the derivative of $\ln Z$ with respect to $|{\bf B}|$,
\beq
M ~=~ -~ \frac{k_B T}{NZ} ~\frac{dZ}{d|{\bf B}|} ~.
\eeq
(We must eventually take the limit $N \rightarrow \infty$).

We should note here that when we actually do the transfer matrix calculations
(on which Figs. 2 - 7 are based), we have not used the assumption made 
in Eqs. (\ref{h1c}) and (\ref{h2c}) that each radical spin is aligned in a 
direction which is determined only by the neighboring cobalt spins. Rather, we
solve for the two eigenvalues of the Hamiltonian of each radical spin which is 
interacting both with its neighboring cobalt spins {\it and} with the applied 
magnetic field. We then take only the larger of these eigenvalues into account
when we integrate out that particular radical spin; the justification for this
is that the two eigenvalues are separated by an energy of order $J$, and the
temperatures of interest are much smaller than $J/k_B$.

While considering the magnetization as a function of the magnetic field, one 
can think of various possible patterns of signs and magnitudes of the 
parameters $\delta \theta_1, \delta \theta_2$ and $\delta \theta_3$. One 
pattern which leads to magnetization plateaus at 0 and $M_S /3$, for a 
magnetic field applied along the $c$ axis, is given by the conditions 
\bea
& & {\rm (i)} ~~\delta \theta_1 + \delta \theta_2 ~,~ \delta \theta_1 + 
\delta \theta_3 ~,~ \delta \theta_2 + \delta \theta_3 ~>~ 0 ~, \non \\
& & {\rm (ii)} ~~\delta \theta_1 ~\ge ~\delta \theta_2 , ~\delta \theta_3 ~,~~
{\rm and}~~ 2 \delta \theta_1 ~>~ \delta \theta_2 + \delta \theta_3 ~.~~~
\label{cond}
\eea
(Condition (i) in Eqs. (\ref{cond}) corresponds to antiferromagnetic 
interactions $J_{i,i+1}$ between neighboring cobalt spins). 
At zero temperature, we then find that there is a magnetization plateau at 
$M=0$ if the strength of the field lies in the range $0 < B < B_1$, where 
\beq
B_1 ~=~ \frac{\sqrt 3}{2} ~\frac{\delta \theta_2 + \delta \theta_3}{g_{eff}} ~
\frac{J}{\mu_B} ~,
\label{b1}
\eeq
a plateau at $M=M_S /3$ if the field lies in the range $B_1 < B < B_2$, where
\beq
B_2 ~=~ \frac{\sqrt 3}{4} ~\frac{2\delta \theta_1 + \delta \theta_2 + \delta 
\theta_3}{g_{eff}} ~\frac{J}{\mu_B} ~,
\label{b2}
\eeq
and a saturation plateau at $M=M_S$ if $B > B_2$. (Note that the condition 
$2 \delta \theta_1 > \delta \theta_2 + \delta \theta_3$ in Eqs. (\ref{cond}) 
is needed in order to have $B_2 > B_1$; otherwise the intermediate plateau at 
$M = M_S /3$ will not exist). As we raise the temperature, the plateaus will 
gradually disappear.

\begin{figure}[t]
\begin{center}
\epsfig{figure=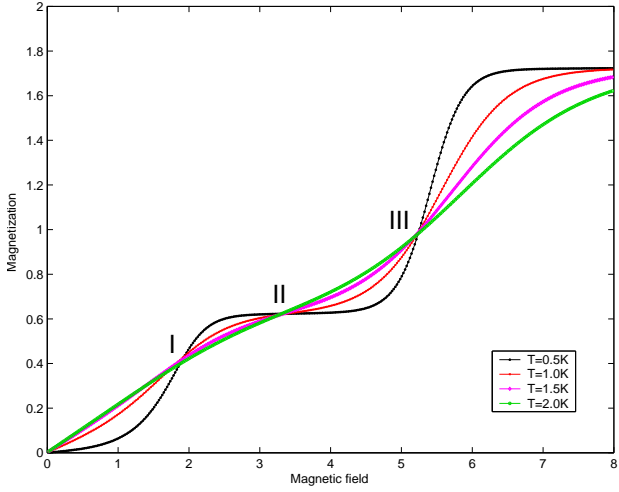,width=8.6cm}
\end{center}
\caption{Magnetization (in units of $\mu_B$) per cobalt-radical pair versus 
the magnetic field (in Tesla) applied along the $c$ axis, for various 
temperatures. The crossing points $I$, $II$ and $III$ are discussed in the
text. (We have taken $J/k_B = 400$ K, $g_C =9$, $g_R = 2$, $\delta 
\theta_1 = \delta \theta_2 = 2.64^o$, and $\delta \theta_3 = - 1.32^o$).}
\end{figure}

\begin{figure}[t]
\begin{center}
\epsfig{figure=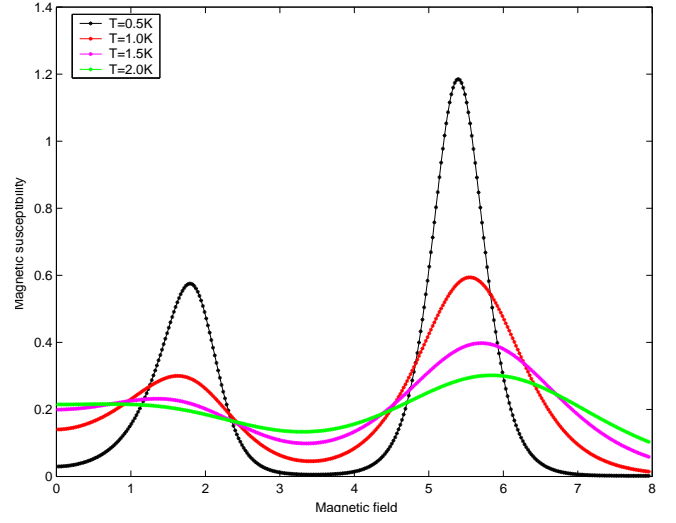,width=8.6cm}
\end{center}
\caption{Magnetic susceptibility (in units of $\mu_B /$ Tesla) per 
cobalt-radical pair versus the magnetic field (in Tesla) applied along the $c$
axis, for various temperatures. ($J/k_B = 400$ K, $g_C =9$, $g_R = 2$, $\delta
\theta_1 = \delta \theta_2 = 2.64^o$, and $\delta \theta_3 = - 1.32^o$).}
\end{figure}

In Fig. 2, we show the magnetization as a function of a magnetic field applied
along the $c$ axis, for one particular choice of the parameters $\delta
\theta_i$ which satisfies the conditions in Eqs. (\ref{cond}), and different
temperatures. For the various parameters given in the caption of Fig. 2 and
using Eqs. (\ref{mag}), (\ref{b1}) and (\ref{b2}), we find that $B_1 = 1.92$ 
Tesla, $B_2 = 4.81$ Tesla, and $M_S /\mu_B = 1.78$. The locations of the 
plateaus in Fig. 2 at the lowest temperature of $0.5$ K are consistent with 
these numbers. We have chosen the parameters $\delta \theta_i$ in such a way 
that the locations of the plateaus and their temperature dependences are in 
rough agreement with the data presented in Ref. \cite{caneschi2}; the agreement
can be improved by changing the value of $g_C$, but we will not do that here.

In Fig. 3, we present the magnetic susceptibility $\chi = (\partial M /\partial
B)_T$ as a function of the magnetic field for different temperatures; these
plots are just given by the derivatives of the plots in Fig. 1. At the lowest
temperature of $0.5$ K, we see peaks at the ends of the magnetization plateaus,
i.e., at $B=B_1$ and $B=B_2$. The peaks get washed out with increasing 
temperature.

We observe three special points labeled $I$, $II$ and $III$ in Fig. 2 where 
the curves for different temperatures seem to cross. In terms of the magnetic 
field (in Tesla) and magnetization (in units of $\mu_B$), these crossing points
lie at $I =(1.93,0.42)$, $II =(3.36,0.62)$, and $III =(5.25,0.99)$. We will 
now provide an explanation of these points based on the transfer matrix method
in the limit of zero temperature.

For $B_1 \le B \le B_2$, we find that there is an exponentially large number 
of degenerate ground states, giving rise to a finite entropy per cobalt-radical
pair at $T=0$. In the limit $N \rightarrow \infty$, the zero temperature 
entropy (in units of $k_B$) per cobalt-radical pair is found to be 
\bea
\frac{S}{k_B} &=& \frac{1}{3} ~\ln ~({\sqrt 2} ~+~ 1) ~~{\rm for}~~ B ~=~ 
B_1 ~, \non \\
&=& \frac{1}{3} ~\ln 2 ~~{\rm for}~~ B_1 ~<~ B ~<~ B_2~, \non \\
&=& \frac{1}{3} ~\ln 3 ~~{\rm for}~~ B ~=~ B_2 ~, \non \\
&=& 0 ~~{\rm for}~~ B ~<~ B_1 ~~{\rm and}~~ B ~>~ B_2 ~.
\label{entropy}
\eea
[We should note here that the numbers given in Eq. (\ref{entropy}) are valid
only for our particular choice of the $\delta \theta_i$, with $\delta \theta_1
= \delta \theta_2 > \delta \theta_3$. If we had chosen $\delta \theta_1 > 
\delta \theta_2 > \delta \theta_3$, the entropy at zero temperature would be
finite only for $B=B_1$ and $B=B_2$].

\begin{figure}[t]
\begin{center}
\epsfig{figure=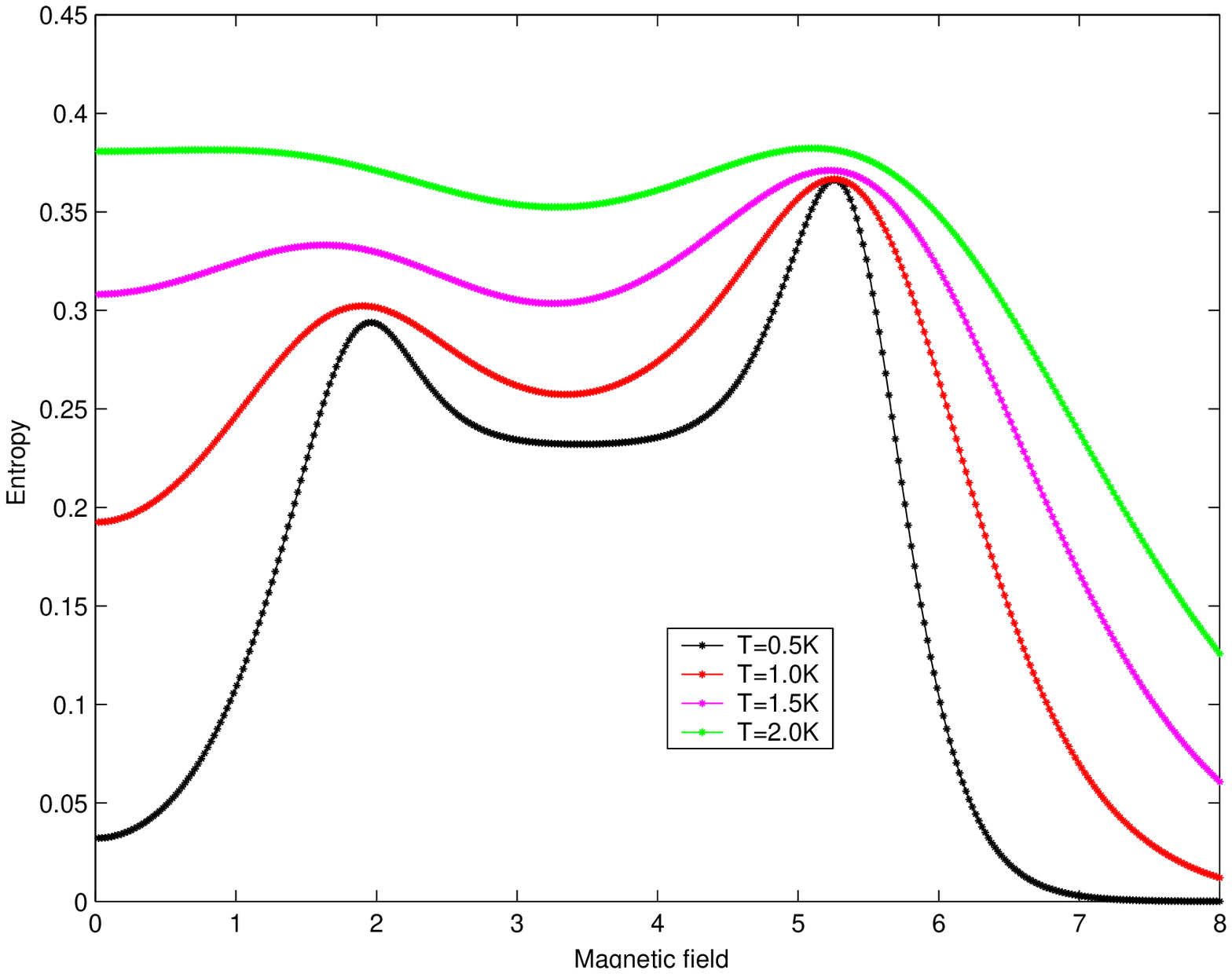,width=8.6cm}
\end{center}
\caption{Entropy (in units of $k_B$) per cobalt-radical pair versus the 
magnetic field (in Tesla) applied along the $c$ axis, for various temperatures.
($J/k_B = 400$ K, $g_C =9$, $g_R = 2$, $\delta \theta_1 = \delta \theta_2 = 
2.64^o$, and $\delta \theta_3 = - 1.32^o$).}
\end{figure}

\begin{figure}[t]
\begin{center}
\epsfig{figure=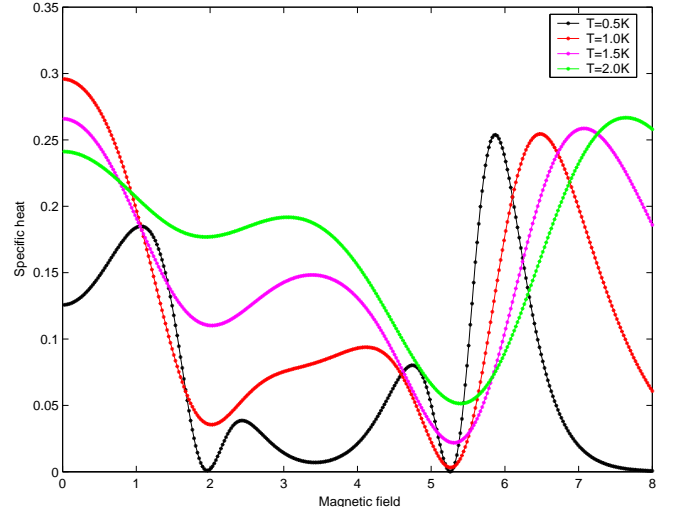,width=8.6cm}
\end{center}
\caption{Specific heat (in units of $k_B$) per cobalt-radical pair versus the 
magnetic field (in Tesla) applied along the $c$ axis, for various temperatures.
($J/k_B = 400$ K, $g_C =9$, $g_R = 2$, $\delta \theta_1 = \delta \theta_2 = 
2.64^o$, and $\delta \theta_3 = - 1.32^o$).}
\end{figure}

The magnetization at $T=0$ is given by an average over all the 
degenerate ground states. For $B=B_1 = 1.92$ Tesla, we find that $M=M_S /(3 
{\sqrt 2}) \simeq 0.42$; this agrees with the location of the first crossing 
point mentioned above. For $B_1 < B < B_2$, we can see why there is a 
degeneracy of $2^{N/3}$; in each group of three successive cobalts, the Ising 
spins $(\sigma_1 , \sigma_2 , \sigma_3)$ can take the orientations $\uparrow 
\downarrow \uparrow$ or $\downarrow \uparrow \uparrow$; different groups of 
three cobalts can take either of these two orientations independently of each 
other. The magnetization of all these states is given by $M_S /3 = 0.59$; this
roughly agrees with the location of the second crossing point; the significance
of the magnetic field value of $3.36$ Tesla at that crossing will be discussed
in the next paragraph. For $B = B_2 = 4.81$ Tesla, the degeneracy of $3^{N/3}$
arises because the Ising spins
in each group of three successive cobalts can independently take the 
orientations $\uparrow \downarrow \uparrow$, $\downarrow \uparrow \uparrow$ or
$\uparrow \uparrow \uparrow$; the average magnetization is therefore given by 
$5M_S /9 = 0.99$ which agrees well with the location of the third crossing 
point (the magnetic field value does not agree so well however). 

We will now see why the second crossing point in Fig. 2 occurs at a magnetic 
field value of $3.36$ Tesla \cite{moessner}. 
We saw above that in each group of three successive cobalts spins $(\sigma_1,
\sigma_2, \sigma_3)$, there are two spin configurations which are degenerate 
in the range $B_1 < B < B_2$, namely, $\uparrow \downarrow \uparrow$ and 
$\downarrow \uparrow \uparrow$. Let us now consider the lowest excitations 
lying above these configurations. There are two kinds of excitations: 
(i) a cobalt spin can flip from down to up, i.e., a group of three cobalts can 
become $\uparrow \uparrow \uparrow$, and (ii) a cobalt spin labeled $\sigma_3$ 
whose neighbors are pointing up can flip from up to down, i.e, a group of six 
cobalts can change from $\downarrow \uparrow \uparrow \uparrow \downarrow 
\uparrow$ to $\downarrow \uparrow \downarrow \uparrow \downarrow \uparrow$. 
The first kind of excitation costs an energy 
\beq
E_+ ~=~ \frac{J}{4} ~(\delta \theta_1 + 2 \delta \theta_2 + \delta \theta_3) ~
-~ \mu_B g_{eff} \frac{B}{\sqrt 3} ~, 
\eeq
and increases the total magnetization by $\mu_B g_{eff} /{\sqrt 3}$. The 
second kind of excitation costs an energy 
\beq
E_- ~=~ -~ \frac{J}{4} ~(\delta \theta_1 + \delta \theta_2 + 2 \delta 
\theta_3) ~+~ \mu_B g_{eff} \frac{B}{\sqrt 3} ~, 
\eeq
and decreases the total magnetization by $\mu_B g_{eff} /{\sqrt 3}$. The
energy costs of the two excitations are equal at a magnetic field given by $B_0
= (B_1 + B_2)/2 = 3.36$ Tesla, where we have used Eqs. (\ref{b1}) and 
(\ref{b2}). At this value of the magnetic field, and at very low temperatures, 
the concentrations of the two kinds of excitations will be small and equal; 
hence the magnetization will lie at its plateau value of $M_S /3$. This 
explains why the different plots in Fig. 2 cross at this value of the magnetic
field and magnetization.

\begin{figure}[t]
\begin{center}
\epsfig{figure=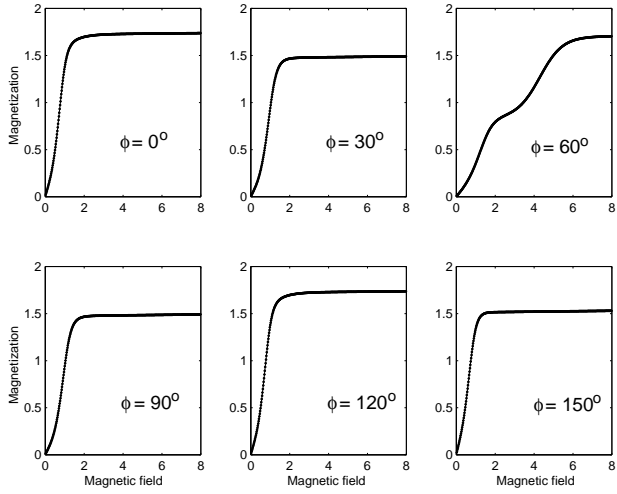,width=8.6cm}
\end{center}
\caption{Magnetization (in units of $\mu_B$) per cobalt-radical pair versus the
magnetic field (in Tesla) applied along six different directions in the $a-b$
plane, for a temperature of $1$ K. ($J/k_B = 400$ K, $g_C =9$, $g_R = 2$, 
$\delta \theta_1 = \delta \theta_2 = 2.64^o$, and $\delta \theta_3 = 
- 1.32^o$).}
\end{figure}

\begin{figure}[t]
\begin{center}
\epsfig{figure=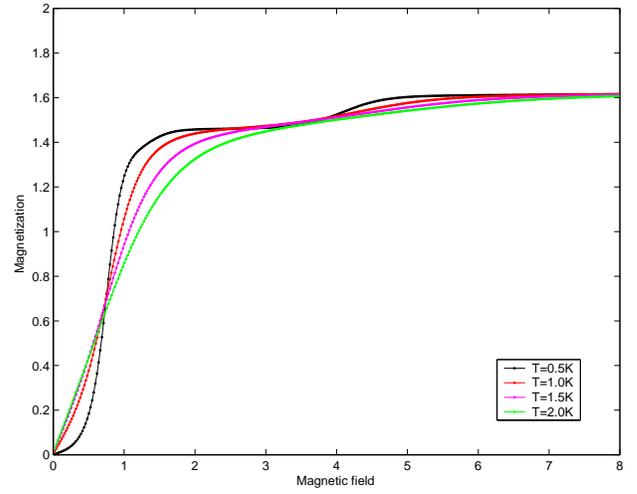,width=8.6cm}
\end{center}
\caption{Magnetization (in units of $\mu_B$) per cobalt-radical pair versus the
magnetic field (in Tesla) averaged over the six different directions shown in 
Fig. 2, for various temperatures. ($J/k_B = 400$ K, $g_C =9$, $g_R = 2$, 
$\delta \theta_1 = \delta \theta_2 = 2.64^o$, and $\delta \theta_3 = 
- 1.32^o$).}
\end{figure}

In Fig. 4, we show the entropy versus the magnetic field for the same values of
parameters and same temperatures as in Fig. 2. We see that at the lowest 
temperature of $0.5$ K, the entropy has a substantial value in the range
$B_1 < B < B_2$, has peaks at $B_1$ and $B_2$, and is quite small for 
$B < B_1$ and $B > B_2$; the values of the entropy at $B_1$ and $B_2$
and on the plateau in between are in agreement with Eq. (\ref{entropy}).
As the temperature is raised, the entropy increases in such a way as
to wash out these features; this is consistent with the disappearance of
the magnetization plateaus in Fig. 2.

Fig. 5 shows the specific heat $C_V = T (\partial S /\partial T)_B$ as a 
function of magnetic field at different temperatures. An interesting feature to
note is that at the lowest temperature of $0.5$ K, the specific heat vanishes
with a parabolic shape at the ends of magnetization plateaus shown in Fig. 1, 
i.e., at $B=B_1$ and $B_2$. This can be understood as 
follows. If $\Delta E$ denotes the energy of a state with respect to the ground
state, the contribution of that state to the specific heat is proportional to
\beq
\frac{C_V}{k_B} ~\sim ~\Bigl( \frac{\Delta E}{k_B T} \Bigr)^2 ~e^{- \Delta E /
k_B T} ~.
\label{cv}
\eeq
At the lowest temperature and at $B=B_1$ and $B_2$, it turns out that all the 
states either have $\Delta E >> k_B T$ (hence their contributions to the 
specific heat are exponentially small and can be ignored), or $\Delta E << k_B
T$. For the latter states, one can show that $\Delta E$ vanishes near $B=B_1$ 
and $B_2$ as $(\mu_B g_{eff} /{\sqrt 3}) |B - B_1|$ and $(\mu_B g_{eff} /
{\sqrt 3}) |B - B_2|$ respectively. From Eq. (\ref{cv}), we see that the 
contributions of these states to the specific heat go as $(B - B_1)^2 /T^2$ 
and $(B - B_2)^2 /T^2$ respectively. This explains the behavior of the 
specific heat in Fig. 5 near $B= B_1$ and $B_2$.

In Fig. 6, we show the magnetization versus the magnetic field applied in the 
$a-b$ plane for six possible directions (parameterized by the angle $\phi$ 
with respect to the projection of ${\bf e}_1$ on that plane), for the same 
values of parameters used in Fig. 2, and a temperature of $1$ K. The six 
directions were chosen with equiangular spacing to cover the full range of 
possible directions from $0^o$ to $180^o$; we recall that the behavior of an 
Ising model does not change if the sign of the magnetic field is reversed, 
i.e., if $\phi \rightarrow \phi + 180^o$. [The projections of the easy axis of
the three cobalts on the $a-b$ plane are given by $0^o$, $120^o$ and $240^o$. 
Since we have chosen $\delta \theta_1 = \delta \theta_2$, we also have a 
symmetry under $\phi \rightarrow 120^o - \phi$. This explains why the plots 
for $\phi = 30^o$ and $90^o$ are identical, as are the plots for $\phi = 0^o$ 
and $120^o$]. We see in Fig. 6 that there is a plateau at intermediate values 
of the magnetization only for a magnetic field direction given by $60^o$; even 
that plateau is much weaker than the plateau seen in Fig. 1 at the same 
temperature. 

Fig. 7 shows the magnetization versus the magnetic field applied in the 
$a-b$ plane, averaged over the six directions indicated in Fig. 6, for various
temperatures. We see that there is no discernible plateau at intermediate 
magnetization even at the lowest temperature of $0.5$ K. This may explain why 
no plateau is observed experimentally when a magnetic field is applied in the 
$a-b$ plane. Since the system consists of several molecular chains, and these 
may happen to be rotated with respect to each other by various amounts in the 
$a-b$ plane, it is possible that the behavior observed experimentally is an 
average of the different directions of the magnetic field in that plane.

Another pattern of signs and magnitudes of the parameters $\delta \theta_1, 
\delta \theta_2$ and $\delta \theta_3$ which leads to magnetization plateaus 
at 0 and $M_S /3$, for a magnetic field applied along the $c$ axis, is 
given by the conditions 
\bea
& & {\rm (i)} ~~\delta \theta_1 + \delta \theta_2 ~>~ 0 ~,~ \delta \theta_1 + 
\delta \theta_3 ~<~ 0 ~,~ \delta \theta_2 + \delta \theta_3 ~<~ 0 ~, \non \\
& & {\rm (ii)} ~~\delta \theta_2 ~\ge ~\delta \theta_1 ~, ~~{\rm and}~~ \delta
\theta_1 + 4 \delta \theta_2 + 3 \delta \theta_3 ~>~ 0 ~.
\eea
We will not discuss the details of this case since the analysis and 
magnetization plots obtained are similar to the case of Eqs. (\ref{cond}) 
considered above.

To summarize, we have studied a model for $CoPhOMe$ in which the tilt
angles of the easy axis of the cobalt spins with respect to the $c$ axis 
vary with period three. We have shown that for certain patterns of these tilt 
angles, the magnetization at low temperatures exhibits plateaus at non-trivial
values if a magnetic field is applied along the $c$ axis, but not if it is
applied in the $a-b$ plane. We have not considered here 
any dynamical effects (arising from the time-dependence of the magnetic field)
for the magnetization; such effects have been discussed earlier for the case 
of a magnetic field applied along the $c$ axis \cite{caneschi2,vindigni}.

\vskip .4 true cm

SR and DS thank R. Sessoli and A. Vindigni for introducing us to this system
and for many stimulating discussions. We also thank A. Vindigni for providing
us in advance with a preliminary version of Ref. \cite{vindigni}. DS thanks R.
Moessner for several useful suggestions and for pointing us to Refs. 5 and 6. 
DS also thanks the Department of Science and Technology, India for financial 
support through grant no. SP/S2/M-11/00.

\end{document}